\documentclass[twocolumn, prb]{revtex4}

\usepackage{graphicx}
\usepackage{dcolumn}
\usepackage{bm}

\begin{document}

\title{ Disordered phase of a two-dimensional Heisenberg Model with $S=1$}

\author {S. Moukouri }

\affiliation{ Department of Physics and  Michigan Center for 
          Theoretical Physics \\
         University of Michigan 2477 Randall Laboratory, Ann Arbor MI 48109}

\begin{abstract}
We study an anisotropic version of the $J_1-J_2$ model with $S=1$. We
find a second order transition from a N\'eel $Q=(\pi,\pi)$ phase to a 
disordered phase with a spin gap. 
\end{abstract}

\maketitle

\section{Introduction}

The Heisenberg model can be mapped to the non-linear sigma model ($NL\sigma$)
with an additional term due to Berry phases. In 1D this term has
a dramatic effect which was first discovered by Haldane \cite{haldane1}. 
Integer spin
 systems have a gap in their excitation spectrum, while half-integer spin
systems are gapless but disordered, with an algebraic decay of spin-spin 
correlations. In the late 1980s it was found that Berry phases do not seem
 to play a role in 2D for pure Heisenberg Hamiltonians \cite{dombre,stone,
zee,haldane2}. Hence,  
all Heisenberg models on the 2D square lattice are on the ordered side of 
the $NL\sigma$ model. This is in agreement with a theorem by Dyson, Lieb, 
and Simon (DLS)\cite{dls}, extended to 2D systems by Neves and 
Peres\cite{neves}, which states 
that for bipartite lattices with $S \ge 1$, the ground state is  
ordered. Furthermore, Monte Carlo simulations \cite{young} 
have convincingly shown that 
the $S=\frac{1}{2}$ model has N\'eel order.

Affleck \cite{affleck}extented to 2D systems a 1D theorem due to Lieb, 
Schultz, and Mattis \cite{lsm} (LSM) which applies to half-integer 
spin systems. He argued that since
this theorem does not apply to integer spin systems, there could be
a difference between the disordered phases of half-integer and integer
spin systems in 2D as well. Haldane \cite{haldane2} argued that if by some 
mechanism it 
was possible to drive the Heisenberg  model into the disordered phase, 
singular topological  contributions known as hedgehogs may be relevant. 
Read and Sachdev \cite{read} carried out a systematic study of the 2D 
Heisenberg model 
in the large-$N$ limit. Their results were in agreement with Affleck and 
Haldane's predictions. They predicted that the nature of the disordered 
phases in 2D is related to the 'spin' value. For odd integer spins, the 
ground state breaks the symmetry of rotation of the lattice by $180$ degres, 
for even integer 'spins' the ground state does not break any lattice symmetry, 
and for half-integer 'spins', the lattice symmetry is broken by $90$ degres.

The usual criticism against such predictions is that they are obtained
in the limit of large spin and they could be invalid for the more
relevant cases of small $S$. It is well possible that a different mechanism
can emerge for small $S$. The $J_1-J_2$ model is the most popular model
in which a possible disordered phase has been searched. For $S=\frac{1}{2}$, 
it is generally accepted that for $0.38 < \frac{J_2}{J_1} < 0.6$, this model  
has a disordered phase. Among the many disordered phases that were
proposed \cite{fradkin},  the columnar dimer phase which was predicted
in Ref.(\cite{read}) seems to gain broader acceptance lately. But, we have 
recently argued \cite{moukouri-TSDMRG3} that this conclusion, which seems 
to be supported by numerical
experiments using exact diagonalization (ED) or series 
expansions \cite{lhuillier}, may be
incorrect. Large scale renormalization group studies on an anisotropic
version of the $J_1-J_2$ model show that in the region were the disordered
phase is expected, physical quantities of the 2D model are nearly 
identical to those of an isolated chain. This suggests that there is
instead a direct transition at $\frac{J_2}{J_1}=0.5$ between the N\'eel
$Q=(\pi,\pi)$ and $Q=(\pi,0)$ phases. At the transition point, 
the system is disordered with algebraic decay of the correlations 
along the chains and exponential decay in the other direction.
This state is consistent with the LSM theorem. 

While the case $S=\frac{1}{2}$ has generated numerous studies \cite{lhuillier},
 other values of $S$ have not been studied to the author's knowledge. Thus,
 the role of topological effects in the $J_1-J_2$ model for small $S$ remains
unknown.  In this letter, we propose to study the case $S=1$. We will apply the 
two-step density-matrix renormalization group 
\cite{moukouri-TSDMRG, moukouri-TSDMRG2} (TSDMRG) to study the 
spatially anisotropic Heisenberg Hamiltonian in 2D,

\begin{eqnarray}
 \nonumber H=J_{\parallel} \sum_{i,l}{\bf S}_{i,l}{\bf S}_{i+1,l}+J_{\perp} \sum_{i,l}{\bf S}_{i,l}{\bf S}_{i,l+1}\\
+J_d \sum_{i,l}({\bf S}_{i,l}{\bf S}_{i+1,l+1}+{\bf S}_{i+1,l}{\bf S}_{i,l+1})
\label{hamiltonian}
\end{eqnarray}

\noindent where $J_{\parallel}$ is the in-chain exchange parameter and is set
 to 1; $J_{\perp}$ and $J_d$ are respectively the transverse and diagonal 
interchain exchanges. Although the Hamiltonian (\ref{hamiltonian}) is 
anisotropic, it retains the basic physics of $J_1-J_2$ model. In the
absence of $J_d$, the ground state is a N\'eel ordered state with 
$Q=(\pi,\pi)$. When $J_d \gg J_{\perp}$, another N\'eel state with
$Q=(\pi,0)$ becomes the ground state. A disordered ground state is
expected in the vicinity of $J_d=\frac{J_{\perp}}{2}$. In this study,
we will only be concerned with the transition from $Q=(\pi,\pi)$ N\'eel
phase to the disordered phase. The lattice size is fixed to $32 \times 33$;
 the transverse coupling is set to $J_{\perp}=0.2$ and $J_d$ is varied from
$J_d=0$ up to the maximally frustrated point $J_d=0.102$, i.e., the point where the
ground state energy is maximal (see Ref.(\cite{moukouri-TSDMRG3})).  We
use periodic boundary conditions (PBC) in the direction of  the chains and open 
boundary conditions (OBC) in the transverse direction. This short paper will
be followed by  a more extensive work \cite{moukouri-TSDMRG4} where a 
finite size analysis is performed.

\section{Method}

 We used the TSDMRG \cite{moukouri-TSDMRG, moukouri-TSDMRG2} to study 
the Hamiltonian (\ref{hamiltonian}). 
The TSDMRG is an extension two 2D anisotropic lattices of
the DMRG method of White \cite{white}. In the first step of the
method, ED or the usual DMRG method are applied to generate a low
energy Hamiltonian of an isolated chain of lenght $L$ keeping $m_1$ states.
 Thus the superblock size is $9 \times {m_1}^2$ for an $S=1$ system.
Then $m_2$ low-lying states of these superblock states, the corresponding
energies, and all the local spin operators are kept. These describe
the renormalized low energy Hamiltonian of a single chain. They form the
starting point of the second step in which $J_{\perp}$ and $J_d$ are switched
on. The coupled chains are studied again by the DMRG method.   
Like the original DMRG method, the TSDMRG is variational. Its convergence depends on 
 $m_1$ and $m_2$, the error is given by $max(\rho_1,\rho_2)$, where
$\rho_1$ and $\rho_2$ are the truncation errors in the first and second steps
respectively.

Since the TSDMRG starts with an isolated chain, a possible criticism of
the method is that it could not effectively couple the chains. This
means that it would eventually miss an ordered magnetic phase.
The source of this criticism is the observation that the DMRG was
introduced to cure the incorrect treatment of the interblock
coupling in the old RG. But this criticism misses the fact that, in the
old RG treatment, dividing the lattice into blocks and treating the
interblock as a perturbation was doomed to failure because both
the intra and inter-block couplings are equal. In the coupled
chain problem, however, when the interchain coupling is small, it
is imperative as Wilson \cite{wilson} put it long ago to separate the 
different energy scales.

\section{Results}

\begin{figure}
\includegraphics[width=3. in, height=2. in]{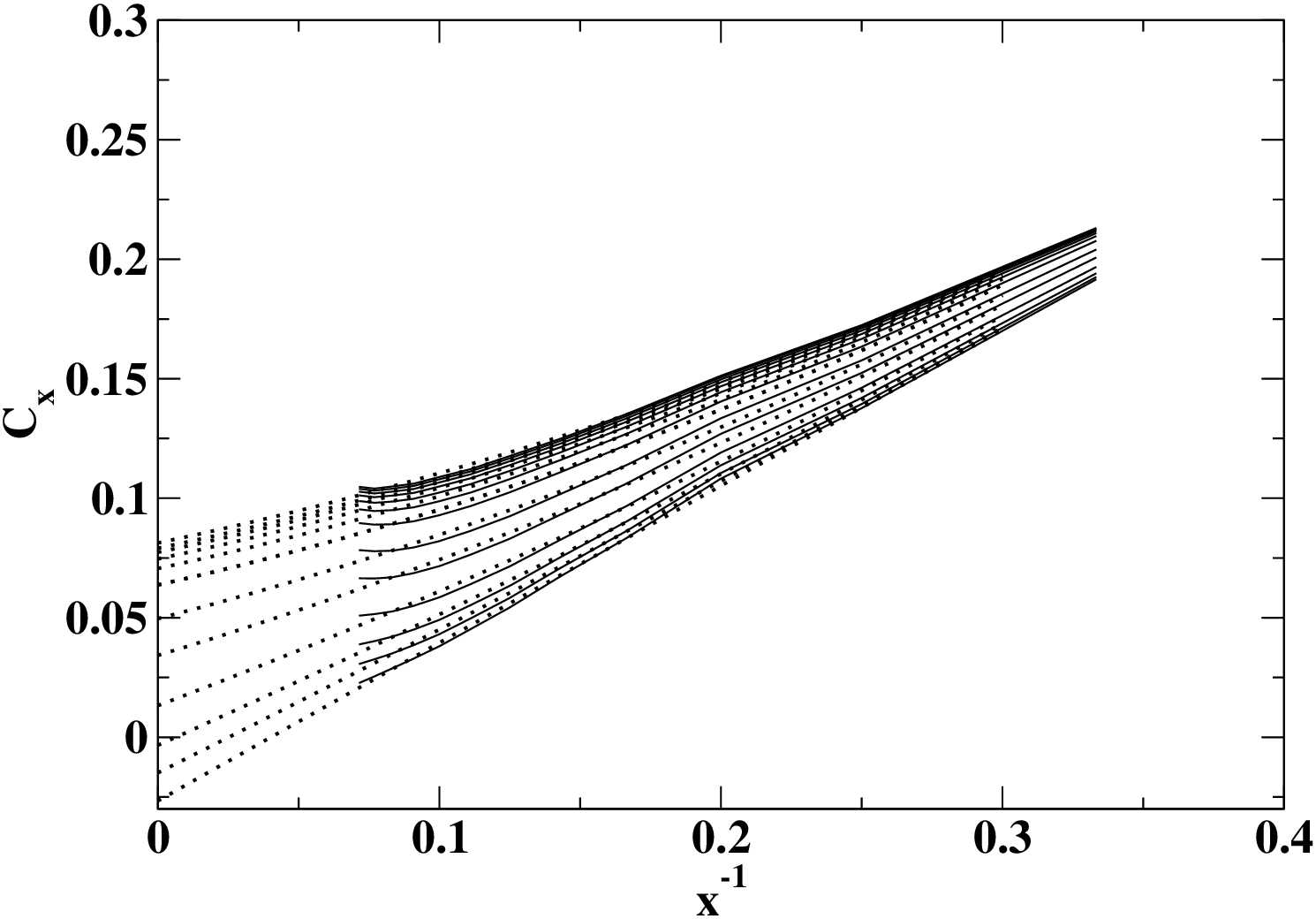}
\caption{Longitudinal spin-spin correlations (full line) and 
their extrapolation (dotted line) for $J_d=0$ (top),
$0.01$, $0.02$, $0.03$, $0.04$, $0.05$, $0.06$, $0.07$, $0.075$, $0.08$,
$0.085$, $0.09$, $0.101$ (bottom) as function of distance.} 
\vspace{0.5cm}
\label{corlpar}
\end{figure}

\begin{figure}
\includegraphics[width=3. in, height=2. in]{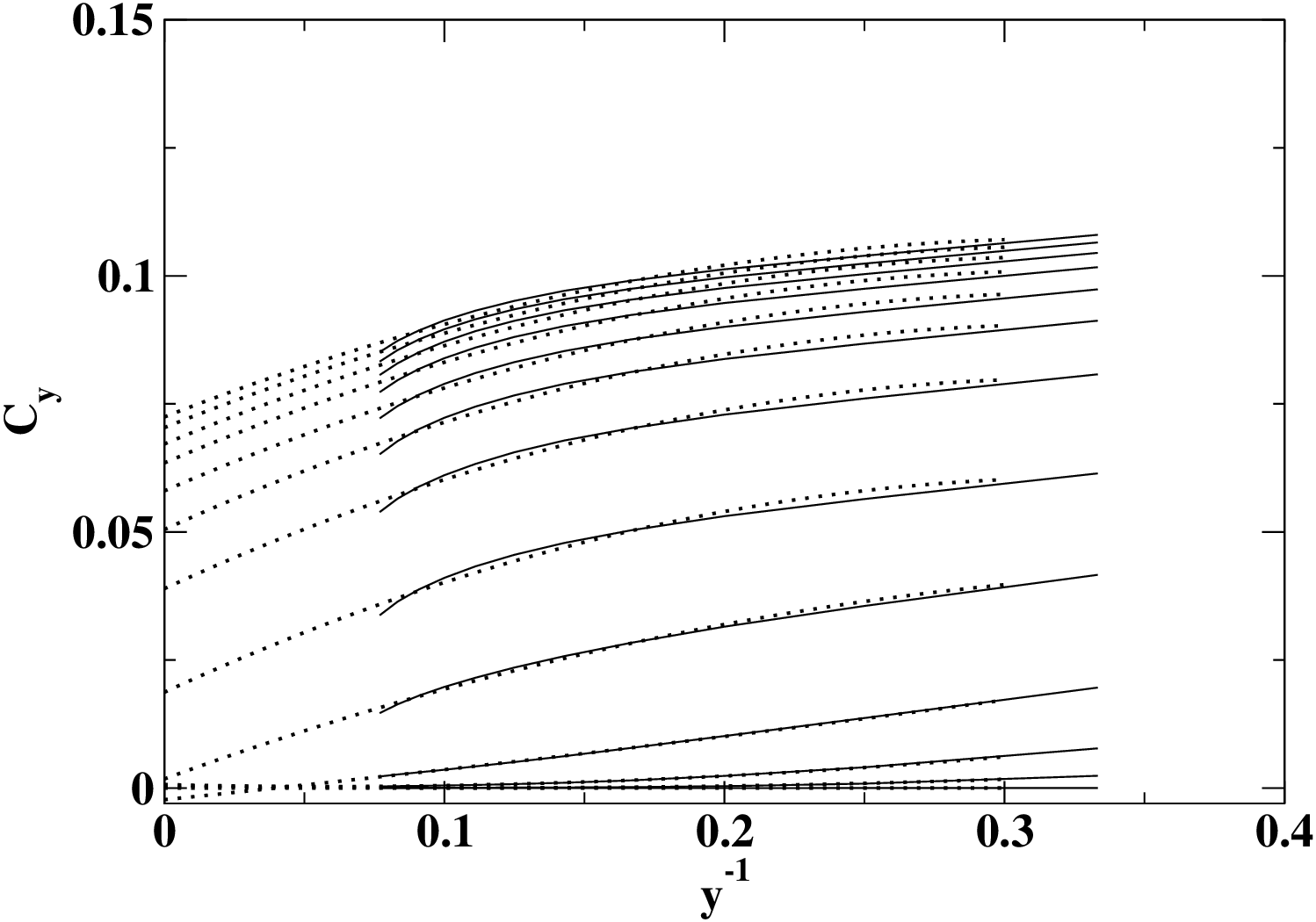}
\caption{Transverse spin-spin correlations (full line) and their 
extrapolation (dotted line) for $J_d=0$ (top),
$0.01$, $0.02$, $0.03$, $0.04$, $0.05$, $0.06$, $0.07$, $0.075$, $0.08$,
$0.085$, $0.09$, $0.101$ (bottom) as function of distance.}
\vspace{0.5cm}
\label{corltran}
\end{figure}

The low energy Hamiltonian for an isolated chain is relatively easy
to obtain, we keep $m_1=162$ states and $L=32$. For this size
the finite size gap is $\Delta=0.4117$ which is very close to
its value in the thermodynamic limit $\Delta_H=0.4105$. This is because
we used PBC and the correlation lenght is about six lattice spacings.
The truncation error during this first step was $\rho_1=5\times 10^{-5}$.
We then kept $m_2=64$ lowest states of the chain to start the second
step. During the second step, the ground state with one of the first
excited triplet states with $S_z=1$ were targeted. The truncation error
during this second step varies from $\rho_2=1.\times 10^{-3}$ in the
magnetic phase to $\rho_2=1.\times 10^{-7}$ in the disordered phase.
This behavior of $\rho_2$ is consistent with previous tests in 
$S=\frac{1}{2}$ systems in Ref.(\cite{alvarez}) where we find that
the accuracy of the TSDMRG increases in the highly frustrated regime.

 We have shown in Ref.(\cite{moukouri-TSDMRG2}) for $S=\frac{1}{2}$
that: (i) the TSDMRG shows a good agreement with QMC for lattices of up to
$32 \times 33$, even if a modest number of states are kept; (ii) spin-spin
correlations extrapolate to a finite value in the thermodynamic limit.
However, because of the strong quantum fluctuations present for 
$S=\frac{1}{2}$, the extrapolated quantities were small and thus could be
doubted. Furthermore, our prediction of a gapless disordered state between
the two magnetic phases has been regarded with a certain skepticism 
\cite{tsvelik, starykh, sindzingre}
because it would be expected that such a state would be unstable
against some relevant perturbation at low energies not reached in our 
simulation. This is not the case for $S=1$, where quantum fluctuations are 
weaker. We thus expect larger extrapolated values $C_{x=\infty}$ and 
$C_{y=\infty}$.
The results in Fig.(\ref{corlpar}) for the correlation function
along the chains,

\begin{equation}
C_x=\frac{1}{3}\langle {\bf S}_{L/2,L/2+1}{\bf S}_{L/2+x,L/2+1} \rangle,
\end{equation}

\noindent and in Fig.(\ref{corltran}) for the correlation function in the 
transverse direction,

\begin{equation} 
C_y=\frac{1}{3}\langle {\bf S}_{L/2,L/2+1} {\bf S}_{L/2,L/2+y} \rangle,
\end{equation}

\noindent unambiguously show that in the weakly frustrated regime, the system is
ordered. Despite the strong anisotropy, $C_{x=\infty}$ and $C_{y=\infty}$
are not very different. The anisotropy is larger   
for small $x$ and $y$. But, due to the difference in the boundary conditions,
$C_x$ seems to reach a plateau while $C_y$ bends downward. This behavior of
$C_y$ is indeed related to the fact that the spins at the edge do not feel
the bulk mean-field created by other chains.

As $J_d$ increases, $C_{x=\infty}$ and $C_{y=\infty}$ decreases and 
vanish at $J_{d_c} \approx 0.085$ and $J_{d_c} \approx 0.075$ respectively.
The difference in the value of $J_{d_c}$  in the two directions is probably 
due to the difference in the boundary conditions. 
  
\begin{figure}
\includegraphics[width=3. in, height=2. in]{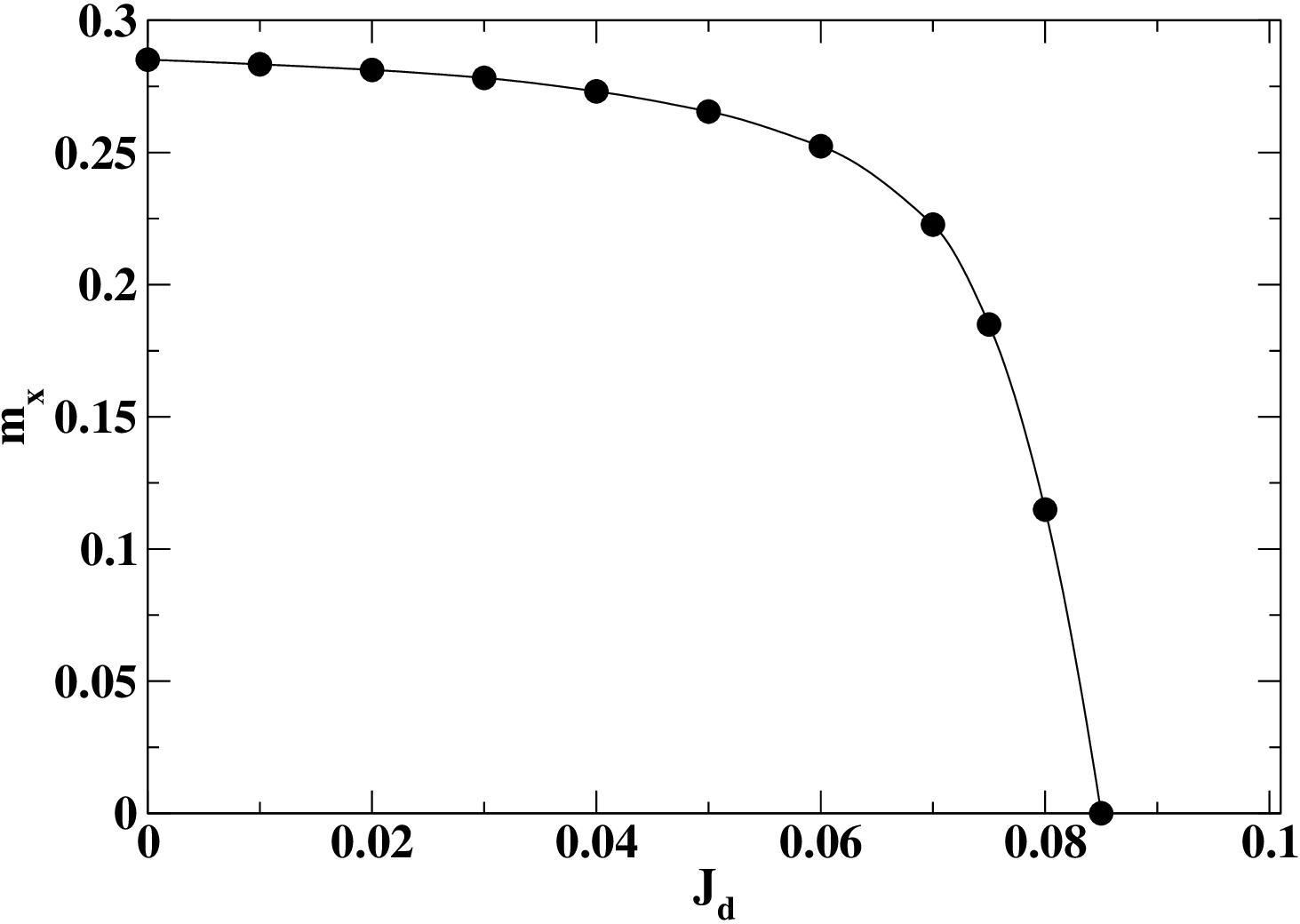}
\caption{Order parameter along the chains as function of $J_d$.} 
\vspace{0.5cm}
\label{orderp}
\end{figure}

\begin{figure}
\includegraphics[width=3. in, height=2. in]{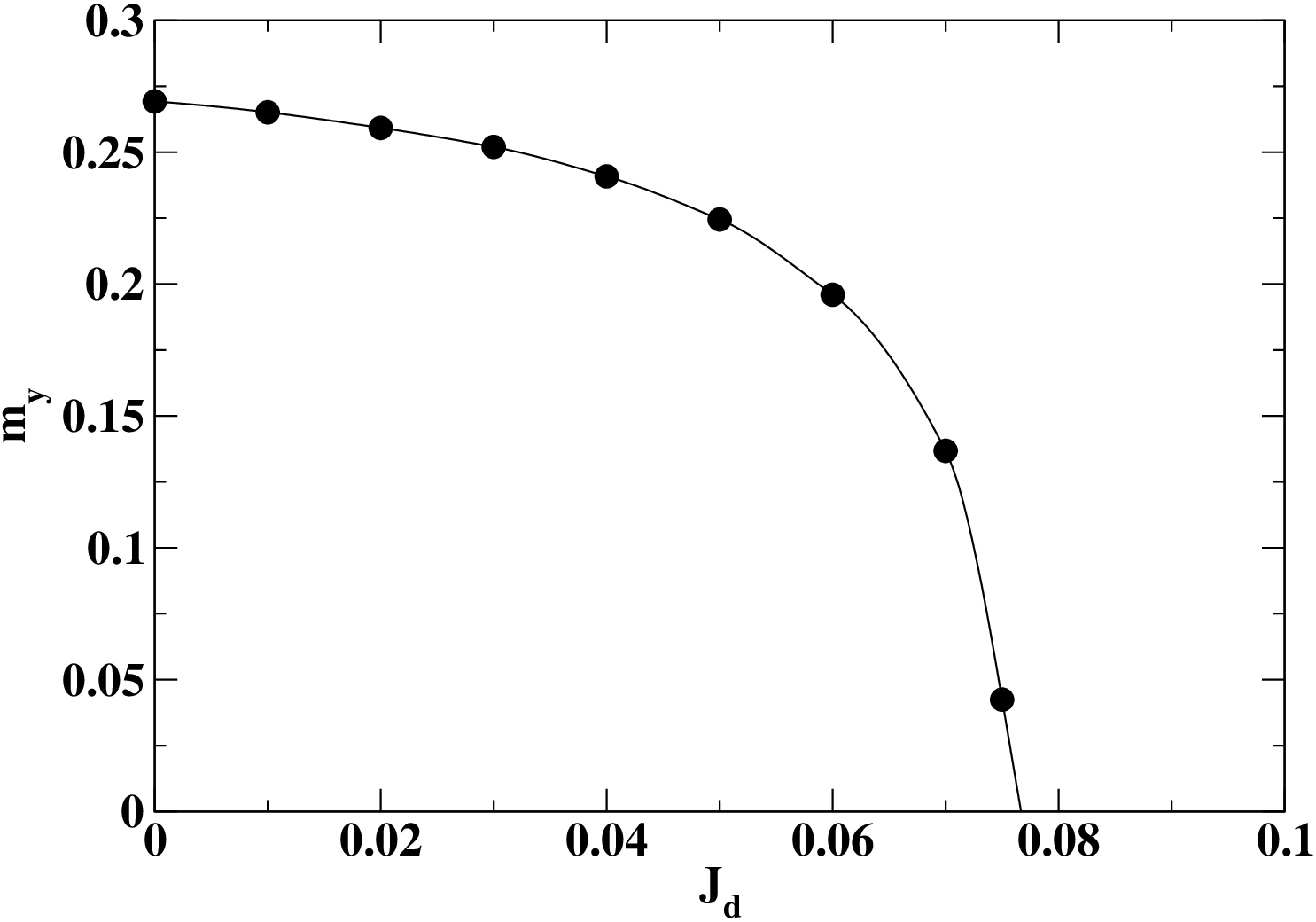}
\caption{Order parameter in the transverse direction as function of $J_d$.}
\vspace{0.5cm}
\label{ordert}
\end{figure}

In Fig.(\ref{orderp}) and Fig.(\ref{ordert}) we plot the corresponding
order parameters $m_x=\sqrt{C_{x=\infty}}$ and 
$m_y=\sqrt{C_{y=\infty}}$. The two curves display the typical
 form of a second order phase transition. However, we have not extracted 
any exponent because even though we believe our results will remain true
in the thermodynamic limit, finite size effects may nevertheless be important 
close to the transition. A systematic analysis of this region is left for
a future study.

\begin{figure}
\includegraphics[width=3. in, height=2. in]{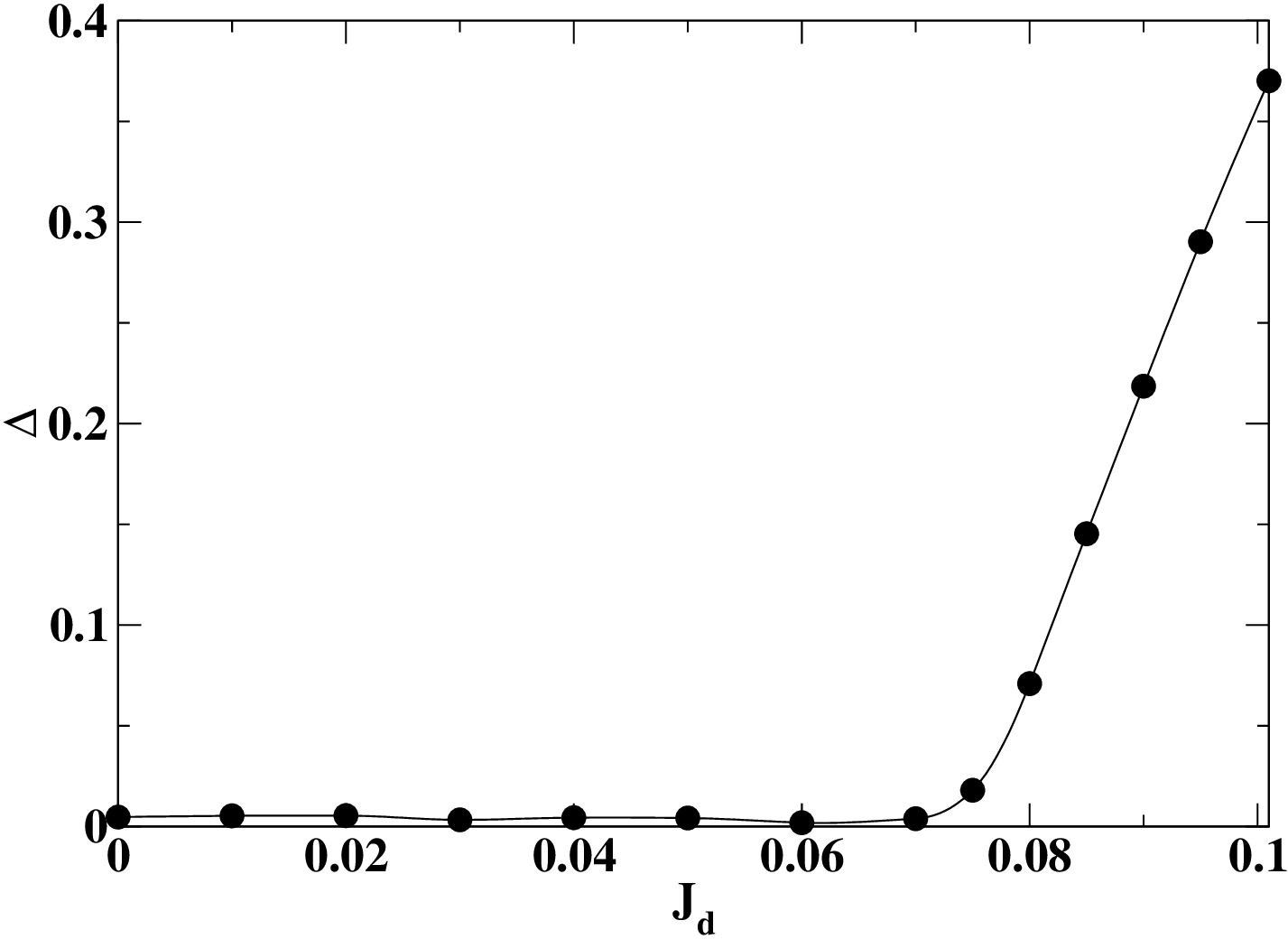}
\caption{Gap as function of $J_d$.}
\vspace{0.5cm}
\label{gap}
\end{figure}

The phase transition is also seen in the spin gap shown in Fig.(\ref{gap}).
 In the ordered state,
the system is gapless. The finite size spin gap is nearly independent
of $J_d$ and is in the order of the truncation error $\rho_2$. At the
transition which occurs at $J_{d_c} \approx 0.075$, $\Delta$ sharply increases
and becomes close to the Haldane gap at the maximally frustrated point
where we find $\Delta=0.3854$ for $J_d=0.102$.  

\section{Conclusion}

In this letter, we have studied an anisotropic version of the $J_1-J_2$
model with $S=1$ using the TSDMRG. We find that for a critical value
of the frustration, the N\'eel ordered phase is destroyed and the system
enters into a disordered  phase with a spin gap. The value of the gap
at the maximally frustrated point is close to that of the Haldane gap
of an isolated chain. This disordered phase is consistent with the large
$N$ prediction of Ref.(\cite{read}). This study shows the striking
difference between integer and half-interger spin systems. For
$S=\frac{1}{2}$, the TSDMRG predicted a direct transition between
the two N\'eel phases with a disordered gapless state at the critical
point. Thus, as we have recently found \cite{moukouri-TSDMRG4}, despite the 
fact that the mechanism of the destruction of the N\'eel phase is independent
of the value of the spins, at the transition point, topological effects
become important leading to the distinction between integer and half-integer
spins.

\begin{acknowledgments}
 We wish to thank K. L. Graham for reading the manuscript. This work was 
supported by the NSF Grant No. DMR-0426775.
\end{acknowledgments}

\end{document}